\documentclass[prb,aps,twocolumn,epsfig,floats,showpacs]{revtex4}
\usepackage{graphicx}
\begin{document}

\title{Dynamic correlated Cu(2) magnetic moments in superconducting
\\YBa$_2$(Cu$_{0.96}$Ni$_{0.04}$)$_3$O$_y$ (y $\sim$ 7)}
\author{J.A. Hodges, P. Bonville, A. Forget}
\affiliation{CEA, Centre d'Etudes de Saclay, DSM/IRAMIS/Service de Physique 
de l'Etat Condens\'e, 
\\ 91191 Gif-sur-Yvette, France}
\author{A. Yaouanc, P. {Dalmas de R\'eotier}}
\affiliation{CEA/DSM/Institut Nanosciences et Cryog\'enie, 38054 Grenoble, 
France}
\author{S.P.~Cottrell}
\affiliation{ISIS Facility, Rutherford Appleton Laboratory, Chilton, Didcot, 
OX11 0QX, UK} 


\begin{abstract}

We have examined the magnetic properties of polycrystalline, superconducting
YBa$_2$(Cu$_{0.96}$Ni$_{0.04}$)$_3$O$_y$ (y $\sim$7, T$_{\rm sc}$ $\sim$75\,K)
using two local probe techniques: $^{170}$Yb M\"ossbauer down to 0.1\,K
and muon spin relaxation ($\mu$SR) down to 1.5\,K. At $0.1 \, {\rm K}$, 
the $^{170}$Yb  measurements show the Cu(2) over essentially
all the sample volume carry magnetically correlated moments which are
static on the time-scale $10^{-9} {\rm s}$. The moments show a distribution 
in size. The correlations are probably short range. 
As the temperature increases, the correlated moments are observed to fluctuate
with measurable rates (in the GHz range) which increase as the temperature 
increases and which show a wide distribution.
The $\mu$SR measurements also evidence that the fluctuation rates increase 
with increasing temperature and there is a distribution. 
The evidenced fluctuating, correlated Cu(2) moments coexist at an atomic level
with superconductivity. 

\end{abstract}

\pacs{74.72.Bk,74.25.Ha,76.80.+y,76.75.+i}

\maketitle

\section {Introduction}

Superconductivity occurs in the cuprates when a sufficient density
of carriers is introduced into the Cu-O planes of the parent compound which is
essentially an antiferromagnetic insulator. 
This feature has stimulated much work directed towards understanding the link 
between magnetism and superconductivity in these compounds. Two particular 
centres of interest concern a), the possible role of spin fluctuations in 
mediating superconductivity and b), the antiferromagnetism -
superconductivity phase diagram as a function of carrier density.
This latter aspect also embraces the question whether or not
antiferromagnetic order and superconductivity are always mutually exclusive on
an atomic level.

In YBa$_2$Cu$_3$O$_y$, the Cu occupy Cu(1) (chain) and Cu(2) (plane) sites.
The phase diagram of extends from $y \simeq$ 6, where the 
compound is a Mott insulator and the Cu(2) of the planes 
order antiferromagnetically to $y \simeq$ 7, where it is an optimally doped 
superconductor (T$_{\rm sc} \sim$ 90\,K) and the Cu(2) do not carry magnetic 
moments. 
The introduction of a low level of doping into the insulator leads to the 
breakdown of the long range order and the introduction of dynamical features. 
Some Cu(2) based magnetic order persists in samples with intermediate doping 
levels which show superconductivity \cite{sidis01}. Moments linked with 
orbital currents have also been evidenced \cite{fauque06} but not confirmed
\cite{MacDougall08}.
In the intermediate doping region, early theoretical studies based on the 
Hubbard and t-J models, envisaged segregation into charge poor 
(antiferromagnetic Cu(2)) and charge rich (superconducting) regions 
\cite{zaanen89,kato90,emery90}, in which case the observed Cu(2) magnetic 
order would involve only part of the sample. Although local probe measurements
on underdoped polycrystalline samples evidence segregation into magnetic and 
non-magnetic regions \cite{hodges91b}, it is problematic, in such samples, to 
experimentally determine whether or not superconductivity and Cu(2)
magnetic order are mutually exclusive at an atomic level. 

The question may thus be asked, are there cases in the cuprates where
the carrier density is high enough for well developed 
superconductivity to exist but where the Cu(2) remain magnetically ordered ?
This is clearly not an omnipresent possibility as shown by the discussions 
concerning the existence of a quantum critical point \cite{sachdev03}, 
but it is still of interest to examine whether Cu(2) magnetic order and 
superconductivity can coexist at an atomic level in specific cases.

Here we show that in absence of any applied magnetic field, all the Cu(2) in 
fully oxidized, superconducting 
YBa$_2$(Cu$_{0.96}$Ni$_{0.04}$)$_3$O$_y$ (y $\sim$7)
carry correlated magnetic moments.
The correlations are probably short-range. 
Nuclear Magnetic Resonance (NMR) results on equivalent
samples have previously shown that field induced staggered Cu(2) magnetic 
moments are nucleated around the substituted Ni both above \cite{bobroff97}
and below \cite{ouazi06} T$_{sc}$.

\section {Samples, Methodology}

For the M\"ossbauer measurements, a single phase polycrystalline sample of 
Y$_{0.975}$Yb$_{0.025}$Ba$_2$(Cu$_{0.96}$Ni$_{0.04}$)$_3$O$_y$
($y \sim$7) was prepared through conventional cycles of 
sintering and crushing, followed by oxygen anneals. The Yb was enriched in the
M\"ossbauer isotope $^{170}$Yb.
We recall that the Yb$^{3+}$ ion is a non-perturbing probe in that it has no 
influence on the superconductivity (neither on T$_{sc}$ nor on the Meissner 
fraction) and it has no influence on any Cu(2) based magnetic order. 
The total signal provided by a M\"ossbauer measurement contains contributions
coming from each and every individual $^{170}$Yb probe in the sample. The
measurements thus provide information which concerns the total sample volume.

An equivalent single phase sample which did not contain any Yb was prepared 
in the same way for the $\mu$SR measurements. 
DC susceptibility measurements, made down to 4.2\,K, provided the 
T$_{\rm sc}$ values of $\sim$75\,K (with or without Yb) which is a typical 
value for a Ni concentration of $4 \,\%$. These measurements also showed 
that as the temperature is reduced, the (negative) Meissner susceptibility 
is essentially independent of temperature below $\sim$50\,K.

The M\"ossbauer measurements on the paramagnetic $^{170}$Yb$^{3+}$ probes were
carried out in absence of any applied field.
We recall the main aspects \cite{hodges91b,vaast97,hodges91a}.
The $^{170}$Yb$^{3+}$ probes, which substitute at the Y$^{3+}$ 
site within the Cu(2) - O bilayers, are randomly distributed throughout 
the sample volume. The average in-plane distance between two probes is 
$\sim$\,5 a or b lattice parameters. The majority of the $^{170}$Yb$^{3+}$ 
thus behave as isolated  probes in that they have no interaction with any 
other of the Yb$^{3+}$. In principle, there will be a spin-spin interaction 
linking the small fraction of the Yb$^{3+}$ which have a Ni as a near 
neighbour but experimentally this interaction plays a negligible role.
The only interaction that is clearly detected by the $^{170}$Yb$^{3+}$ is the 
molecular field which acts on its spin. This field is present 
when the nearest neighbour Cu(2) to a Yb$^{3+}$ carry correlated magnetic 
moments. These magnetically correlated Cu(2) are initially nucleated around 
each Ni. 
If we observe that all (or a fraction) of the probes experience a 
molecular field, this shows the Cu(2) in all (or in a fraction)
of the sample carry magnetic moments that are correlated. This local probe 
technique provides no information concerning the size of the Cu(2) moments
nor their magnetic correlation length. However, it provides information
concerning the fluctuation rate of the molecular field (the fluctuation rate 
of the correlated Cu(2) magnetic moments) provided this falls within the 
$^{170}$Yb M\"ossbauer frequency window for which the lower bound is 
$\sim$10$^9$\,s$^{-1}$.

An introduction to the positive muon spin rotation/relaxation ($\mu$SR) 
techniques is given in Refs. \onlinecite{Dalmas97,Dalmas04}. 
The measurements were carried out at the MuSR spectrometer of the 
ISIS facility, Rutherford-Appleton Laboratory, Chilton, UK. 

Both the substituted $^{170}$Yb$^{3+}$ probes and the implanted muon probes
furnish information concerning the local magnetic properties of the matrix by 
detecting the internal fields that are produced on them. 
For the $^{170}$Yb$^{3+}$ probe, it is a molecular field that is detected, 
whereas for the muon probe, it is the longer range dipolar field. The two 
techniques thus examine the local magnetic properties over somewhat different 
length scales. Both techniques can also provide information concerning the 
fluctuation rates of the field.

\section {Experimental results}

\subsection {$^{170}$Yb M\"ossbauer probe data and analysis}
\label{Mossbauer}

The $^{170}$Yb M\"ossbauer measurements 
(I$_g$ = 0, I$_{ex}$ = 2, E$_{\gamma}$ = 84\,keV, 1\,mm/s = 68\,MHz) were 
made using a source of Tm$^{\star}$B$_{12}$ and a linear velocity sweep.

The analysis of the experimental lineshapes to be presented below, is made in 
terms of the electro-nuclear (Breit-Rabi) Hamiltonian comprising a hyperfine
interaction and an electronic Zeeman term :
\begin{eqnarray}
 {\cal H} = S' . {\bf A} . I + \mu_{\rm B} S'. {\bf g} . H(t). 
\label{eqn1}
\end{eqnarray}
{\bf A} and {\bf g} are the known hyperfine and g-tensors associated with the 
Yb$^{3+}$ ground state Kramers doublet \cite{hodges91a}, $\mu_{\rm B}$ is the
Bohr magneton and H(t) the fluctuating molecular field acting on the Yb$^{3+}$
which has an effective spin $S'$=1/2. The energy equivalent of each of the two
terms is of order 0.1\,K.

\begin{figure}
\includegraphics[width=0.45\textwidth]{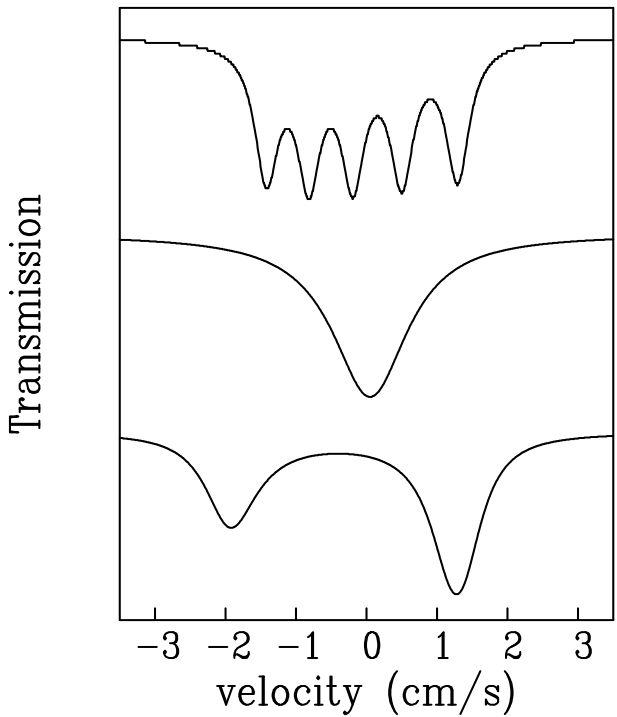}
\caption{Calculated $^{170}$Yb M\"ossbauer absorption spectra corresponding to
Hamiltonian (1) for a field H(t) of $ 0.2 \, {\rm T}$ fluctuating at rates 
below (top), within (middle) and above (bottom) the accessible frequency 
window.  See the text for details. These calculated spectra correspond to the 
forms of the (sub)spectra evidenced experimentally in 
Fig.\ref{figybmossdata}.}
\label{figybmosssim}
\end{figure}

We first show on Fig.\ref{figybmosssim} simulated lineshapes corresponding to 
Eq.~\ref{eqn1} for different fluctuation rates of the molecular field. 
The three different lineshapes correspond to the three types of 
(sub)spectra that are encountered experimentally. The field intensity is fixed
at the value $0.2 \, {\rm T}$. 
The top lineshape (multicomponent lineshape) was calculated from 
Eq.~\ref{eqn1} with the field ``static'' on the 
$^{170}$Yb M\"ossbauer frequency scale, $\it{i.e}$ if the field fluctuates, it
does so at a rate which is below the lower limit of the M\"ossbauer frequency 
window ($\sim$10$^{9}$s$^{-1}$). The middle lineshape (single component 
lineshape) was obtained with the field fluctuating within
the frequency window at a rate 5.$\times$10$^{10}$s$^{-1}$ and the bottom 
lineshape (doublet lineshape) was calculated with the field  
fluctuating above the frequency window (fluctuating too fast to be detected). 
This lineshape also corresponds to the case when the field is simply absent. 
Experimentally, the assessment as to whether, at a particular temperature, the
field has no influence on the lineshape because it fluctuates too fast or
because it is absent, has to be made by examining the thermal dependence of 
the field in the temperature range where it is visible.

Experimental data were obtained over the range 0.1 to 70\,K and data for three
selected temperatures are shown on Fig.\ref{figybmossdata}. 
We first note that although only one type of sub-spectrum (the multicomponent 
lineshape but with two slightly different forms) is present at 
0.1\,K, at higher temperatures quite different subspectra are simultaneously
present. For example, on Fig.\ref{figybmossdata} at 2.5\,K, the three 
subspectra corresponding to the three examples on Fig.\ref{figybmosssim} are
all present with different relative weights. Our local probe technique thus 
shows that at specific temperatures, it is possible to identify the 
coexistence of quite different local behaviours. The same situation was also 
evidenced in YBa$_2$(Cu$_{1-x}$Co$_{x}$)$_3$O$_7$ \cite{vaast97}. 

At 0.1\,K, only the multicomponent lineshape shown at the top of 
Fig.\ref{figybmosssim} is present. This indicates that each of the Yb$^{3+}$ 
spins throughout the sample experiences a molecular field which is static on 
the time scale 10$^{-9}$s. The presence of this field on all of the 
probes indicates that essentially all the Cu(2) of the sample carry
magnetic moments which are at least short range correlated and which appear 
static. 
The best data fit is obtained with a molecular field that shows a distribution
in size. The effect of the distribution is well mimicked by fitting in terms 
of two fields (0.22\,T on 62\% of the probes and 0.07\,T on 38\%) 
(Fig.{\ref{figybmossdata} top). The weighted mean provides the average field 
(0.16\,T) and the difference between the two fields (0.15\,T) provides an 
estimate of the range of the distribution.
Our observation of a distribution in the size of the molecular field suggests 
there is a distribution in the size of the correlated Cu(2) magnetic moments. 
$^{17}$O NMR measurements on analogous Ni substituted samples 
(Sec.~\ref{Discussion}) have evidenced that the size of the
staggered Cu(2) magnetic moments observed in an applied magnetic field
depend on the distance between a Cu(2) and a Ni. Our results evidence that 
correlated Cu(2) moments exist in absence of an applied field and they show 
an intrinsic distribution in size.
The average size of the molecular field obtained here is similar to 
that observed on Yb$^{3+}$ in superconducting 
YBa$_2$(Cu$_{0.96}$Co$_{0.04}$)$_3$O$_7$ \cite{vaast97}, where the Cu(2) 
moments which give rise to the field have an average value of a fraction of a
$\mu_B$ \cite{hodges02}. We thus suggest the average size of the correlated 
Cu(2) moments in the present Ni substituted samples is of the order of a 
fraction of a $\mu_B$.
We have no direct information concerning the directional properties of the 
correlated Cu(2) magnetic moments. However, since we find the average size and
the average direction of the molecular field observed here are similar to 
those observed on the $^{170}$Yb probe in other cuprates where the Cu(2) 
moments lie in the (ab) plane \cite{vaast97,hodges02}, it is possible that 
here also, the Cu(2) magnetic moments also lie towards the (ab) plane. 

\begin{figure}
\includegraphics[width=0.45\textwidth]{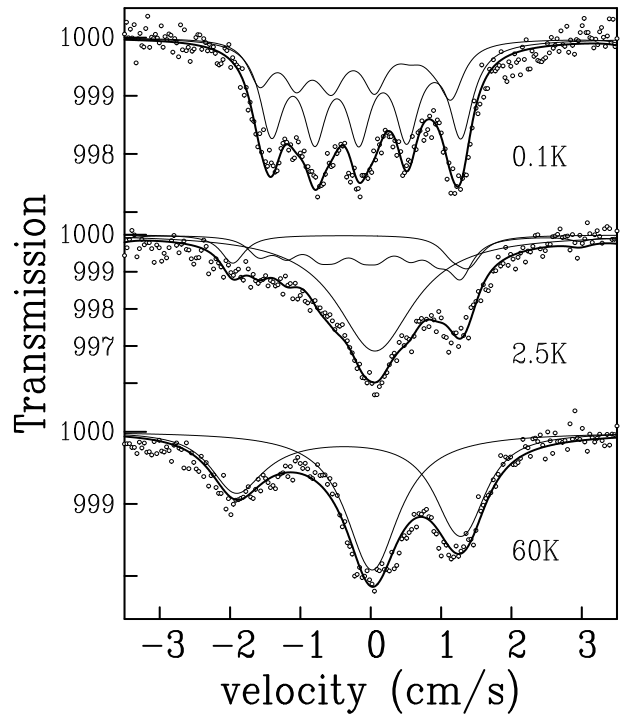}
\caption{$^{170}$Yb M\"ossbauer absorption spectra for
Y$_{0.975}$Yb$_{0.025}$Ba$_2$(Cu$_{0.96}$Ni$_{0.04}$)$_3$O$_y$ ($y \sim$ 7). 
Different subspectra are present and the line fits are explained in the text.}
\label{figybmossdata}
\end{figure}

The variation of the line shapes as a function of increasing temperature 
follows  roughly the same evolution observed in 
YBa$_2$(Cu$_{0.96}$Co$_{0.04}$)$_3$O$_y$ (y $\sim$7) \cite{vaast97} but the 
changes take place at much lower temperatures in the sample substituted 
with Ni than in the sample substituted with Co.
In YBa$_2$(Cu$_{0.96}$Ni$_{0.04}$)$_3$O$_7$ 
there is a relatively rapid change up to $\sim$3.0\,K}:  
at 2.5\,K (Fig.\ref{figybmossdata}), a ``static'' molecular field 
(multicomponent subspectrum) is now present on only $\sim$\,30\% of the 
Yb$^{3+}$  ions and no field (doublet subspectrum) is visible on 
$\sim$\,12\% of the Yb$^{3+}$ with the remainder corresponding to the 
singlet component lineshape. Above $\sim$\,3.0\,K, the subspectrum 
corresponding to a ``static'' field is no longer visible and there is a 
progressive change in the relative weights of 
the two remaining subspectra. Fig.\ref{figybmossdata} shows 
that at 60\,K, the relative weight of the subspectrum corresponding to no 
visible field (doublet subspectrum) has increased to $\sim$\,50\%. The 
analysis of the lineshapes at the different temperatures evidences 
a) at each particular temperature there is a wide distribution in the local 
fluctuation rates and b) the average fluctuation rate increases as the 
temperature is increased. 
Because of the wide distribution in the rates, it is difficult to obtain 
accurate values for the average rates. We simply 
estimate that on increasing the temperature to 60\,K, this average rate 
increases progressively by two to three orders of magnitude above the 
threshold value of 10$^9$\,s$^{-1}$.

If the evolution in YBa$_2$(Cu$_{0.96}$Ni$_{0.04}$)$_3$O$_7$ is considered as
a function of decreasing temperature, the fluctuation rates decrease 
progressively. It thus seems likely that the correlated moments continue to 
fluctuate at 0.1\,K and below, but with rates that are below 10$^9$\,s$^{-1}$
and thus too low to be experimentally accessible using $^{170}$Yb.

\subsection {$\mu$SR probe data and analysis}
\label{muons}

$\mu$SR measurements have been carried out on 
YBa$_2$(Cu$_{1-x}$Ni$_{x}$)$_3$O$_y$ by 
Bucci {\it et al.} \cite{bucci94} with the aim of examining the 
influence of the Ni on the magnetic penetration depth. The measurements
were made down to 35\,K and did not incidentally evidence the 
influence of the magnetic fluctuations on the asymmetry.
As shown below, this influence is only visible well below 35\,K.

\begin{figure}
\includegraphics[width=0.45\textwidth]{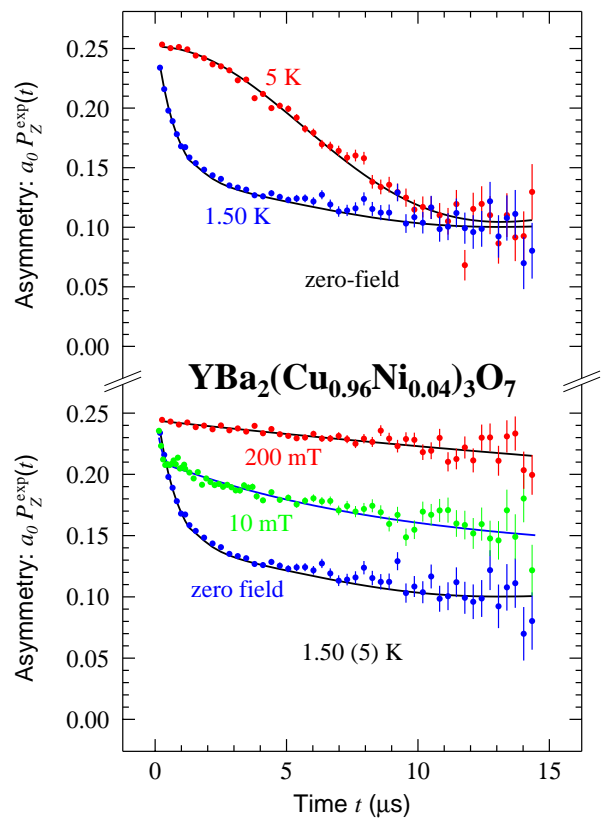}
\caption{(color online). Examples of $\mu$SR asymmetries versus time for
YBa$_2$(Cu$_{0.96}$Ni$_{0.04}$)$_3$O$_y$ ($y \sim$ 7). The asymmetries are 
independent of temperature down to $\sim$5\,K.
The change that occurs between $5$ and $1.5 \, {\rm K}$ (upper panel)
is attributed to a change in the electron based magnetism. The strong 
dependence on applied longitudinal field (lower panel) points to the 
``quasi-static'' nature of the electron based magnetism. 
The line adjustments are described in the text.} 
 \label{fignimuon}
\end{figure}

Our $\mu$SR measurements were carried out from 
$100 \, {\rm K}$ to $1.5 \, {\rm K}$ in zero applied field and at 1.5\,K in 
longitudinal fields (applied in the field cooled configuration) up to 200\,mT.
In zero applied field, the measured asymmetry, {\sl i.e.} the $\mu$SR signal,
is essentially independent of temperature
from 100 to $\sim$5\,K and it then changes progressively as the temperature
is further lowered (Fig.~\ref{fignimuon}). 

The measured spectra are expressed as the product of $a_{\rm 0}$, the  
effective asymmetry of the muon decay, and $P_Z^{\rm exp}(t)$, the 
polarisation function of interest \cite{Dalmas97}.
$a_{\rm 0}P_Z^{\rm exp}(t)$ is a sum of two 
components: the first originating from the sample and the second  
from the sample holder and surroundings. We write
\begin{eqnarray}
a_{\rm 0}P_Z^{\rm exp}(t) =  a_{\rm s} P_Z^{\rm s}(t) + a_{\rm bg},
\label{eqn2}
\end{eqnarray}
where $a_{\rm bg} = 0.099$.  

From 100 to $\sim$ 5\,K, the asymmetry is nicely modelled by the Kubo-Toyabe
function, {\sl i.e.} $ P_Z^{\rm s}(t) = P_Z^{\rm KT}(t)$, linked with the 
interaction between the nuclear magnetic moments and the muon spin. We find 
$a_{\rm s} = 0.153 \, (1)$  and for the field width at the muon site 
$\Delta^{KT} =  0.146 \, (2) \, {\rm mT}$. 
This $\Delta^{KT}$ value is close to that for unsubstituted 
YBa$_2$Cu$_3$O$_7$ \cite{nishida88}.

As the temperature is lowered below $\sim$ 5\,K, the time dependence of the
asymmetry becomes progressively more rapid. This behaviour is due to the
influence of dipolar fields associated with electron based magnetic moments.

The asymmetries below $\sim$5\,K are not well described using the 
product of $P_Z^{\rm KT}(t)$ and a simple exponential function 
$\exp(-\lambda_Z t)$. 
We found they are well accounted for either by taking the 
product with a stretched exponential $\exp[-(\lambda_Z t)^{\beta}]$ or by 
using a two component model: 
\begin{eqnarray}
a_{\rm s} P_Z^{\rm s}(t) 
=  a_{\rm 1} P_Z^{\rm KT}(t) \exp \left(-\lambda_Z t \right)
+ a_{\rm 2} P_Z^{\rm KT}(t).
\label{eqn3}
\end{eqnarray}
where the electron based magnetism influences the 
asymmetry in part of the sample (of relative volume $a_1/a_{\rm s}$) 
and does not influence the asymmetry in the remaining part (of relative volume
$a_2/a_{\rm s}$). This component model is physically equivalent to the 
component model used to interpret the $^{170}$Yb data 
(section \ref{Mossbauer}).

With the stretched exponential model, we find that $\beta$ remains below 1.0
and $\lambda_Z$ increases with decreasing temperature. These results 
indicate a), there is a distribution in the spin-lattice fluctuation rates 
(which we relate to the distribution in the fluctuation rate of the 
magnetically correlated Cu(2) evidenced in section \ref{Mossbauer})
and b), the average electron based magnetic fluctuation rate decreases with 
decreasing temperature (as also evidenced in section \ref{Mossbauer}).

With the two component model, at 3.0, 2.2, and 1.5\,K respectively, we find  
$a_1/a_{\rm s}$, the magnetic fraction, amounts to 0.28(16), 0.42(6) and 
0.78(10) and $\lambda_Z$, the muon spin relaxation rate, amounts to 
0.22(13)\,$\times$\,10$^6$, 0.52(9)\,$\times$\,10$^6$ and 
1.29(5)\,$\times$\,10$^6$ s$^{-1}$ respectively. 
According to this approach, as the temperature is progressively
lowered a), an increasing fraction of the sample experiences electronic based 
magnetic fluctuations which have slowed down to enter the $\mu$SR frequency 
window and b), the average electron-based magnetic fluctuation rate decreases.
We anticipate that $\mu$SR measurements at low enough temperatures 
below 1.5\,K when analysed with the two component model will show that the 
asymmetries of all the muons are influenced by interactions with electron 
based magnetic moments.

The $^{170}$Yb and $\mu$SR analyses thus lead to a common general description
in terms of correlated Cu(2) magnetic moments with temperature dependent 
fluctuation rates. In addition, it seems likely that the approximate $30\%$ 
volume fraction of the sample where the molecular field is ``static'' on the 
$^{170}$Yb time-scale at $2.5 \, {\rm K}$, corresponds to the volume fraction 
($\sim28\,\%$ at $3.0 \, {\rm K}$ , $\sim 42\,\%$ at $2.2 \, {\rm K}$)   
where $\mu$SR evidences electron-based magnetic moments.
The observation that at temperatures just above 5\,K, the fluctuations
are too fast to be detected by $\mu$SR, whereas they are detected by 
$^{170}$Yb indicates that the frequency window accessible for $\mu$SR is lower
than that for $^{170}$Yb.

To further examine the Cu(2) correlations, at $1.5 \, {\rm K}$ we have 
measured the 
influence of applied longitudinal magnetic fields of 10 and
200\,mT (Fig.~\ref{fignimuon}). In these fields, the nuclear moments 
no longer influence the asymmetry. The strong dependence on applied field 
points to the ``quasi-static'' nature of the electron-based magnetic moments.

The fact that a field as small as 
$B_{\rm ext} = 10 \, {\rm mT}$ has a strong influence on the 
asymmetry means that some of the correlations are characterised by a 
fluctuation rate smaller than 
$\gamma_\mu B_{\rm ext} \simeq 10 ^{7} \, {s}^{-1}$ 
(the muon gyromagnetic ratio $\gamma_\mu$ = 851.615 Mrad s$^{-1}$ T$^{-1}$). 
These particular fluctuations are too slow to influence the $^{170}$Yb 
lineshapes and they can be linked to the ``static'' subspectrum of 
Figs.\ref{figybmosssim} and \ref{figybmossdata}.
In fact, the form of the asymmetry with $B_{\rm ext} = 10 \, {\rm mT}$ 
is relatively complex. At short times 
it is suggestive of an overdamped oscillation which would be indicative of 
correlations over lengths of a few lattice spacings \cite{Yaouanc08}.
However, because we know from the $^{170}$Yb and the $\mu$SR measurements 
carried out in zero applied field that distributions exist in both the size 
and fluctuation rates of the correlated Cu(2) magnetic moments, it is not 
possible to carry out a productive quantitative analysis. This is also the 
case for the results obtained with $B_{\rm ext} = 200 \, {\rm mT}$, where the 
asymmetry has an exponential form with 
$\lambda_Z = 0.021 \, (1) \, \mu{\rm s}^{-1}$ and where the whole asymmetry 
is accounted for. 

$\mu$SR measurements have shown that when Ni is substituted into the different
superconducting system La$_{2-x}$Sr$_x$CuO$_4$, the dynamics of the Cu spin 
correlations which develop depend both on temperature (the fluctuation 
rates of the correlated Cu decrease with decreasing temperature) and on Ni
content \cite{adachi08}. 

\section{Discussion}
\label{Discussion}

The $^{170}$Yb M\"ossbauer and $\mu$SR analyses corroborate each other. Both 
analyses support the view that when the Cu in YBa$_2$Cu$_3$O$_y$ is 
substituted by 4\% Ni, correlated and fluctuating Cu(2) magnetic moments 
(having a distribution in their size and in their fluctuation rate) are 
present over essentially all the whole sample volume.
This suggests the length scale around a Ni over which the Cu(2) carry  
correlated magnetic moments is of the order of a few a/b lattice constants. 
The length scale around a Ni over which the Cu(2) carry staggered
paramagnatic moments is similar \cite{ouazi06}. 

Since essentially all the Cu(2) carry correlated magnetic moments and the 
sample remains superconducting, the compound contains both 
localised holes (Cu(2) magnetic moments) and delocalised holes (which lead to 
superconductivity). 
We recall that we have no precise information concerning the size of the Cu(2)
moments (section \ref{Mossbauer})
and consequently the density of the localised holes is not known. 
In addition, it is difficult to assess the level of the superconducting 
condensate in the planes. Since the relation between T$_{\rm sc}$ and the 
condensate density may be non-linear in well doped samples \cite{bernhard95}, 
the fact that T$_{\rm sc}$ remains as high as 75\,K does not automatically 
entail that the plane condensate density approaches that in unsubstituted
optimally doped YBa$_2$Cu$_3$O$_7$.
However, even though the condensate density in the chains may also contribute 
to maintaining superconductivity in the YBa$_2$Cu$_3$O$_y$ \cite{bernhard95}, 
it seems very unlikely that T$_{\rm sc}$ could be as high as 75\,K unless 
there is a significant contribution from a condensate density in the planes.

When substituted in YBa$_2$Cu$_3$O$_7$, Ni enters both the chain and plane 
sites with a significant fraction entering the plane site \cite{adachi00}. 
The structure remains orthorhombic and there is essentially no change in the 
oxygen level nor in the doping level 
\cite{clayhold89a,clayhold89b,bobroff97,pimenov05,bringley88,mirza94}. 
Penetration depth measurements, made down to 1.5\,K, show that the 
superconducting condensate density increases progressively as the temperature 
is lowered \cite{bonn94}.
Above T$_{\rm sc}$, the samples remain metallic and local spin susceptibility 
measurements \cite{dupree92} provide no evidence of a pseudogap.
However, optical conductivity measurements indicates that a gap opens in the 
c-axis conductivity\cite{pimenov05}.
The substitution of Ni also introduces paramagnetic moments which are 
compatible with effective spins of 1/2 to 1 \cite{mirza94,mendels94}. 
In the sibling compound Bi$_2$Sr$_2$CaCu$_2$O$_{8+\delta}$, the substitution 
of Ni does not affect the superconducting gap \cite{hudson01}. 

Nuclear Quadrupole Resonance (NQR) measurements on superconducting  
YBa$_2$Cu$_3$O$_y$ substituted with Ni made above T$_{sc}$ evidence that 
1/T$_1$, the nuclear spin relaxation rate, increases as the concentration of 
Ni increases \cite{tokunaga97}. This indicates that the Cu(2) AF spin 
fluctuations that govern 1/T$_1$ progressively change with Ni content.

Superconducting samples of YBa$_2$(Cu$_{1-x}$Ni$_{x}$)$_3$O$_y$
have been examined both above and below T$_{\rm sc}$ using NMR in 6\,T on 
$^{17}$O substituted in the Cu(2)-O planes \cite{bobroff97,ouazi06}. 
The observed hyperfine coupling is dominated by coupling to the spin 
polarisation of the two nearest in-plane Cu(2) \cite{yoshinari90,takigawa89}. 
The measurements evidence a field induced staggered polarisation of the Cu(2) 
moments around each Ni and 
provide a direct static signature of the magnetic correlations within the 
Cu(2)-O planes. The $^{17}$O probe is situated within a Cu(2)-O plane whereas
the $^{170}$Yb probe is situated between the two Cu(2)-O planes of a bilayer. 
With this probe, we evidence that Cu(2) magnetic correlations exist in 
absence of any applied field and they extend over the bilayers. 

A possible mechanism through which substituted impurity spins may induce Cu(2)
moments in the 
normal state of {\it{underdoped}} cuprates (spin-gap phase) has been examined
theoretically within the t-J model treated with RVB mean-field theory
\cite{kilian99}. The sea of spinons which couple to the localised impurity
spin is polarised by an external field and leads to a staggered Cu(2) spin
polarisation. The polarisation decreases as r$^{-3}$ with distance from the Ni
impurity and the correlation length could be adjusted to reproduce the 
observed $^{17}$O NMR line broadening \cite{bobroff97}. This treatment 
pertains to underdoped cuprates where there is a pseudo-gap in the spin 
excitation spectrum. It does not appear to be relevant to the present case 
where there is no evidence of a pseudo-gap from local susceptibility 
measurements \cite{dupree92} (a gap is however evidenced in the c-axis 
conductivity \cite{pimenov05}) and where there is no applied magnetic field.
Theoretical studies have also shown that d-wave superconductivity and 
antiferromagnetic order and $\pi$ triplet pairs can exist near half filling
\cite{murakani98}. 
In addition, local antiferromagnetic order may appear 
near impurities and near some surfaces in a d-wave superconductor 
\cite{ohashi99} and a phenomenological description based on 
Ginzburg-Landau theory, has suggested that induced antiferromagnetic moments 
may be nucleated in superconducting samples such that there are spatially 
varying order parameters \cite{kohno99}. 
The appropriate theoretical description of the omnipresent
nature of the correlated Cu(2) moments in well doped Ni substituted 
superconducting samples in zero applied field remains to be obtained.

To date, spontaneous correlated magnetic Cu(2) moments have been evidenced in 
fully oxidised superconducting samples of YBa$_2$Cu$_3$O$_7$ substituted with
both Ni (this work) and with Co or Fe \cite{vaast97,hodges02}. 
Further information concerning the properties of magnetically correlated 
Cu(2) in Co substituted single crystals is given in the following
paper \cite{hodges09b}. Whereas each type 
of substitution (Ni or Co/Fe) lowers the superconducting transition 
temperature by approximately the same amount, each introduces quite different 
changes in some of the other properties. For example, Co enters only the 
Cu(1) site, the sample becomes underdoped and there is a pseudo-gap, 
seen for example in the local susceptibility \cite{dupree92}. In 
contrast, Ni enters both the Cu(1) and the Cu(2) sites, the carrier density
is not lowered and local susceptibility measurements show no evidence of a 
pseudo-gap \cite{dupree92}.
Consequently there is no univocal link between the appearance of magnetically
correlated Cu(2) and the site occupied by the substituted cation nor with the
fact that the substitution lowers or does not lower the doping level. There is
no link either with the presence or absence of a spin susceptibility
pseudo-gap.

We note that when a non-magnetic ion, for example Zn, is substituted
into fully oxidised YBa$_2$Cu$_3$O$_7$, neither muon probe measurements 
\cite{garcia-munoz91} nor our $^{170}$Yb measurements (made down to 1.4\,K,
unpublished) provide any evidence of magnetically correlated Cu(2) moments. 
NMR measurements do however evidence field induced paramagnetic moments 
\cite{ouazi06}. 

A straightforward feature thus appears to link the particular substituting 
cation and the correlated Cu(2) moments : these are observed when the 
substituting cation carries an intrinsic magnetic moment, irrespective both 
of the site it occupies and of the other changes it produces.

\section{Conclusions}

The present local probe measurements show that correlated Cu(2) 
magnetic moments are present over essentially all the sample volume of fully 
oxydised, optimally doped, superconducting 
YBa$_2$(Cu$_{0.96}$Ni$_{0.04}$)$_3$O$_y$ (y $\sim$7).
The moments show distributions in their sizes and in their fluctuation rates
which fall typically in the GHz range. The average fluctuation rate increases
as the temperature increases. The Cu(2) moments, whose size, direction and 
correlation length remain to be established, coexist on an atomic level 
with high temperature superconductivity. The Cu(2)-O network of the planes 
is capable of supporting superconductivity when all the Cu(2) carry
fluctuating correlated magnetic moments.

\section{Acknowledgements}

J.A.H thanks Julien Bobroff for useful discussions and Nadine Genand-Riondet 
for assistance.

\end{document}